\documentclass[journal=ancac3,manuscript=article]{achemso}  
\usepackage{amsmath}
\usepackage[T1]{fontenc}
\usepackage{mciteplus}
\usepackage[hidelinks,colorlinks=true,citecolor=magenta]{hyperref}

\author{Sung-Hoon Lee} \email{lsh@khu.ac.kr}
\affiliation{Department of Applied Physics, Kyung Hee University, Yongin 17104, Republic of Korea}

\author{Doohee Cho} \email{dooheecho@yonsei.ac.kr}
\affiliation{Department of Physics, Yonsei University, Seoul 03722, Republic of Korea}

\title{Surface Reconstruction-Driven Insulating Behavior in Metallic Charge-Density-Wave 1T-TaSe$_\mathbf{2}$}

\begin{document}

\begin{abstract}

Bulk 1T-TaSe$_2$ is metallic, yet its surface consistently exhibits an insulating gap---a dichotomy long attributed to a surface Mott insulator driven by enhanced electron correlations. Using density functional theory calculations, we show that this insulating surface instead originates from a charge-density-wave (CDW) stacking reconstruction. Whereas the bulk stabilizes a single-layer CDW stacking that supports metallic transport, the surface energetically favors a bilayer stacking, in which interlayer hybridization of Ta $5d_{z^2}$ orbitals opens a $\sim$0.4~eV gap---a band insulator requiring no on-site Coulomb repulsion. This reconstruction is the thermodynamic ground state for slab thicknesses from two to eight layers, and the calculated surface density of states quantitatively reproduces scanning tunneling spectra of both insulating and metallic domains. CDW surface reconstruction, rather than Mott physics, thus governs the surface electronic structure of 1T-TaSe$_2$, providing a unified explanation for the observed coexistence of metallic and insulating domains.

\end{abstract}

\maketitle

\newpage

\section*{Introduction}

The layered compound 1T-TaSe$_2$ hosts one of the most puzzling surface phenomena in charge-density-wave (CDW) materials: although the bulk is unambiguously metallic \cite{disalvo1974, wilson1969}, scanning tunneling spectroscopy (STS) \cite{colonna2005, chen2022b, zhang2022a, tian2024} and angle-resolved photoemission spectroscopy (ARPES) \cite{perfetti2003, sayers2020, ren2025, straub2025, mignani2025} consistently reveal insulating domains at the surface, with a gap of $\sim$0.5~eV coexisting with metallic regions. The same insulating character persists in ultrathin samples down to the bilayer limit \cite{tian2024}. For two decades, this behavior has been attributed to a ``surface Mott insulator''---a scenario in which the topmost CDW layer decouples from the metallic bulk and becomes a correlated insulator \cite{colonna2005, perfetti2003, sayers2020, chen2022b, zhang2022a, tian2024, ren2025}. In this work, we suggest that this long-standing interpretation may not be appropriate.

The key to our argument lies in the vertical stacking of the CDW. In the commensurate CDW phase, both 1T-TaS$_2$ and 1T-TaSe$_2$ form $\sqrt{13}\times\sqrt{13}$ ``Star-of-David'' (SoD) clusters of thirteen Ta atoms, each contributing one unpaired electron in a narrow, half-filled Ta $5d_{z^2}$ band \cite{wilson1974, fazekas1979, wilson1975, rossnagel2011}. A single layer of either material is a Mott insulator \cite{lin2020, chen2020, nakata2021, vano2021}. However, the bulk properties are dictated not by in-plane correlations alone but by how these CDW clusters stack along the out-of-plane direction \cite{darancet2014, ritschel2015}. In 1T-TaS$_2$, the bulk adopts a bilayer (``\textit{AL}'') stacking in which vertically aligned SoD centers dimerize across the interface, splitting the half-filled band into bonding and antibonding states and producing a band insulator rather than a Mott insulator \cite{ritschel2018, lee2019, tanda1984, *nakanishi1984, naito1986, ganal1990, ishiguro1991, vonwitte2019, stahl2020, wang2020}. In 1T-TaSe$_2$, by contrast, the bulk prefers a single-layer (``\textit{L}'') stacking with laterally shifted SoD centers \cite{moncton1976, brouwer1980, naito1985, wiegers2001, wang2023}, which yields oblique interlayer coupling and a metallic band structure---consistent with the observed metallic transport \cite{disalvo1974, wilson1969}.

Recent studies on 1T-TaS$_2$ have shown that surface termination critically determines the nature of the gap: a bilayer-terminated surface hosts a hybridization gap, whereas a single-layer termination produces a Mott gap \cite{butler2020, wu2022, petocchi2022, lee2023,yang2024}. Moreover, the single-layer-terminated surface spontaneously reconstructs to adopt the bilayer configuration that minimizes surface energy \cite{lee2023}. These findings prompted us to ask whether a similar reconstruction could occur at the surface of 1T-TaSe$_2$---and whether it could explain the insulating surface without invoking Mott physics.

Here, using density functional theory calculations, we demonstrate that the answers to both questions are affirmative. We find that the surface of 1T-TaSe$_2$ reconstructs from the bulk-preferred \textit{L} stacking to an \textit{A}-interface bilayer, in which vertically aligned SoD centers hybridize to open a $\sim$0.4~eV gap through a purely single-particle mechanism. This reconstructed surface is the thermodynamic ground state for slab thicknesses from two to eight layers and quantitatively reproduces the experimentally measured STS spectra for both insulating and metallic domains. The insulating surface of 1T-TaSe$_2$ is therefore a band insulator arising from CDW surface reconstruction, not a Mott insulator.

\section*{Results}

\subsection*{CDW stacking energetics and band structures of bulk 1T-TaSe$_\mathbf{2}$}

We first establish the stacking-dependent properties of bulk 1T-TaSe$_2$ that underpin the surface analysis. Following the notation of Ref.~\cite{lee2019}, the relative lateral displacement of SoD centers between adjacent layers defines five symmetry-distinct interface types---\textit{A}, \textit{B}, \textit{C}, \textit{M}, and \textit{L}---determined by which of the 13 Ta sites in the SoD cluster of one layer lies beneath the center of the adjacent layer (Figure~1a,b). Single-layer stacking repeats one interface type throughout the crystal; bilayer stacking alternates a vertically aligned \textit{A} interface with a shifted interface (e.g., \textit{AL}).

Figure~1c--e compares the calculated total energy, van der Waals energy, and interlayer spacing for all stacking configurations in both 1T-TaSe$_2$ and 1T-TaS$_2$. In both materials, the \textit{L} and \textit{AL} configurations are far more stable than the other configurations and are separated from them by 39--75~meV per SoD cluster (meV/SoD). The stability of the \textit{L} interface originates from two factors: (i) reduced interlayer spacing (up to 1.46\% smaller than the \textit{A} interface in TaSe$_2$; Figure~1e), which enhances van der Waals attraction by 220~meV/SoD (Figure~1d), and (ii) interlayer Se--Se $\pi$-type bonding between chalcogen $p_z$ orbitals that bridge the SoD centers of adjacent layers (Figure~S1). The \textit{C} interface, despite a similarly small spacing, lacks this bonding channel (Figure~S1) and is the least stable configuration (Figure~1c).

A subtle but crucial distinction between the two materials lies in the competition between \textit{L} and \textit{AL}: in 1T-TaSe$_2$, \textit{L} is favored by 8~meV/SoD, whereas in 1T-TaS$_2$, \textit{AL} is favored by 6~meV/SoD. This reversal, governed by the differing extent of chalcogen bonding, accounts for the experimentally observed single-layer stacking in TaSe$_2$ \cite{moncton1976, brouwer1980, naito1985, wiegers2001} versus bilayer stacking in TaS$_2$ \cite{tanda1984, *nakanishi1984, naito1986, ganal1990, ishiguro1991, vonwitte2019, stahl2020, wang2020}. We note that experimental studies typically report a stacking shift of $2\mathbf{a}$ without distinguishing the sign. Our calculations reveal a pronounced difference between the $+2\mathbf{a}$ (\textit{C} interface) and $-2\mathbf{a}$ (\textit{L} interface) configurations, with the \textit{L} interface being far more stable---a distinction that has not been resolved experimentally but is critical for understanding the surface reconstruction.

The electronic consequences of stacking are dramatic (Figure~1f). The \textit{A} stacking produces a half-filled 1D band along $\Gamma$--A with a bandwidth of $\sim$0.9~eV---approximately twice that of 1T-TaS$_2$ \cite{darancet2014} due to stronger Se--Se coupling. The ground-state \textit{L} stacking redirects this coupling obliquely (\(\mathbf{T_s} = -2\mathbf{a} + \mathbf{c}\)), reducing the $k_z$ bandwidth by half while broadening the in-plane dispersion, yielding a metallic band structure consistent with transport measurements \cite{disalvo1974, wilson1969}. The \textit{AL} stacking---the bulk ground state of 1T-TaS$_2$---folds and gaps the 1D band through Peierls dimerization \cite{lee2019} (Figure~S2): the alternating interlayer bonds double the period along $c$, folding the half-filled band along $\Gamma$--A into the halved Brillouin zone and opening a $\sim$0.1~eV gap at the new zone boundary. This stacking-dependent metal--insulator contrast in the bulk sets the stage for understanding the surface.

\subsection*{Surface CDW reconstruction and the insulating surface}
We now turn to the central question: what happens at the surface? To address this, we model the surface using an 8-layer slab, systematically varying the topmost stacking interface while fixing the bottom of the slab in the bulk ground-state \textit{L} stacking.

We begin with the bulk-terminated surface (\textit{LLLLLLL}, Figure~2a), which is metallic, as expected. The Ta $5d_{z^2}$ spectral weight at the SoD center of the top layer (red shading) contributes strongly to bands at the Fermi level, and the projected density of states (PDOS) for the top three layers exhibits no gap.

Replacing the topmost interface with the \textit{A} type (\textit{LLLLLLA}, Figure~2b) transforms the surface from metallic to insulating. The top-layer PDOS now exhibits two sharp peaks at approximately $\pm$0.2~eV, separated by a gap of $\sim$0.4~eV. This gap arises from interlayer hybridization: the vertically aligned SoD centers of the top two layers split the half-filled $5d_{z^2}$ band into bonding and antibonding states. A small residual density of states at the Fermi level, arising from coupling to the metallic third layer, vanishes entirely when the next interface also adopts the \textit{A} type (\textit{LLLLALA}, Figure~2c). Other surface terminations (\textit{B}, \textit{C}, \textit{M}) remain metallic, with only the \textit{B} and \textit{M} terminations showing a shallow pseudogap due to intermediate coupling strength (Figure~S3).

The insulating \textit{LLLLLLA} surface is not merely a possible configuration: it is the thermodynamic ground state. The calculated surface formation energy of the bulk-terminated \textit{LLLLLLL} surface is 2.23~eV/SoD (for surfaces, energies per SoD cluster are equivalent to energies per $\sqrt{13}\times\sqrt{13}$ surface unit cell), and replacing the topmost \textit{L} interface with an \textit{A} interface lowers this energy by $\sim$20~meV/SoD (Figure~2d). The \textit{LLLLALA} configuration, in which the reconstruction extends one layer deeper, is only $\sim$3~meV/SoD higher than \textit{LLLLLLL}, close enough in energy to also appear on the surface. Surfaces terminated by \textit{B}, \textit{M}, or \textit{C} interfaces are significantly less stable. The surface of 1T-TaSe$_2$ therefore has a strong thermodynamic driving force to reconstruct from the metallic bulk stacking into an insulating bilayer, analogous to the CDW surface reconstruction in 1T-TaS$_2$ \cite{lee2023}, albeit starting from a different initial condition. The energetic origin of this reversal is simple: because the topmost layer lacks an upper neighbor, the van der Waals penalty of the \textit{A} interface is reduced and the electronic gain from its dimerization is enhanced at the surface, tipping the delicate bulk balance in favor of \textit{A} (Figure~S4).

The calculated PDOS for all Ta atoms in the surface layer (Figure~2e) directly connects these predictions with experiment. The insulating surfaces (\textit{LLLLLLA}, \textit{LLLLALA}) reproduce the $dI/dV$ spectrum measured on insulating domains of bulk 1T-TaSe$_2$ (Figure~2f), matching both the gap magnitude and the asymmetric peak structure; this agreement is insensitive to $U$, the spectrum remaining nearly unchanged at $U=0.5$~eV (Figure~S5). The metallic surface (\textit{LLLLLLL}) matches the $dI/dV$ spectrum of the metallic domains. This one-to-one correspondence identifies insulating domains as reconstructed \textit{A}-interface bilayer surfaces, and metallic domains as regions retaining the bulk \textit{L} stacking.

To explore the role of the stacking beneath the surface bilayer, we further computed all 13 symmetry-inequivalent choices of the second-topmost interface \textit{X} (\textit{LLLLXA}), as well as all 13 topmost interfaces above a buried \textit{A} interface (\textit{LLLLAX}), using 7-layer slabs (Figure~S6). Only three subsurface interfaces are energetically competitive---\textit{L} (the ground state), \textit{I} ($+7.6$), and \textit{H} ($+9.6$~meV/SoD)---and these nearly degenerate variants share the same $\sim$0.4~eV gap while differing measurably in the empty-state peak intensity and gap-edge profile (Figure~S6b), explaining why nominally identical \textit{A}-terminated surfaces can exhibit distinct STS spectra. In contrast, every buried-\textit{A} configuration is highly unstable ($\geq$75~meV/SoD above the ground state): the \textit{A} bilayer is stabilized exclusively at the outermost interface; the reconstruction is strictly a surface phenomenon.

\subsection*{Thickness dependence}

The surface reconstruction is robust across all sample thicknesses. Figure~3 compares the slab energies for uniform \textit{L} stacking and bilayer-reconstructed configurations from 2L to 8L. The energy reference ($\Delta E_\text{slab} = 0$) corresponds to the thick-slab limit of the ground-state configuration, in which both surfaces are terminated by \textit{A}-interface bilayers (i.e., \textit{ALLLLLA}). At every thickness, the configuration with an \textit{A}-interface bilayer at the surface is energetically favored over uniform \textit{L} stacking. The energy gain is largest for the thinnest slab ($\sim$70~meV/SoD for 2L) and converges to $\sim$20~meV/SoD in the thick-slab limit, consistent with the surface energy difference in Figure~2d.

This energetic stability is directly reflected in the electronic structure. Figure~4a shows the top-layer PDOS for the ground-state configuration at each thickness: the characteristic double-peak structure with a $\sim$0.4~eV gap persists from 2L through 8L. Such thickness-independent behavior is a natural consequence of the band-insulator mechanism---the gap originates from local bilayer hybridization, not from collective phenomena that would be sensitive to the total number of layers. Consistent with this picture, the calculated spectra at each thickness show excellent agreement with STS measurements on 1T-TaSe$_2$ samples of corresponding thickness \cite{tian2024} (Figure~4b).

\section*{Discussion}

Our results challenge the two-decade-old ``surface Mott insulator'' interpretation of 1T-TaSe$_2$ \cite{colonna2005, perfetti2003, sayers2020, chen2022b, zhang2022a, tian2024, ren2025} and instead support a CDW-driven surface reconstruction mechanism. While both scenarios involve the opening of an energy gap, their underlying origins are fundamentally different: a Mott insulator arises from strong electron correlations that localize charge, whereas the band insulator identified here emerges from single-particle interlayer hybridization. We now discuss the robustness of this conclusion, its experimental implications, and its broader significance.

\textit{Role of electron correlations and the choice of $U=0$.} In 1T-TaS$_2$, a moderate $U=1.25$~eV was needed to capture the Mott physics of the surface single layer \cite{lee2023}. In 1T-TaSe$_2$, stronger Se--Se interlayer bonding produces wider bandwidths and weaker effective correlations, making $U=0$ an appropriate choice. This conclusion is supported at multiple levels: the calculated bulk ground state (\textit{L} stacking) matches experimental observations; the surface PDOS quantitatively reproduces STS data (Figures~2f and 4b); and the $U$-dependent bulk phase diagram (Figure~S7) provides an independent constraint: increasing $U$ drives the bulk ground state from \textit{L} to \textit{AL} stacking, with the crossover at $U\approx 0.76$~eV. Since 1T-TaSe$_2$ experimentally adopts \textit{L} stacking, $U$ must be below this threshold, and $U=0$ is the most consistent value. Even at $U=0$---where the bulk correctly favors the metallic \textit{L} stacking---the surface bilayer reconstruction is already energetically preferred, demonstrating that the insulating surface is an intrinsic consequence of interlayer hybridization, not an artifact of an overestimated $U$.

Importantly, $U=0$ does not mean correlations are absent, since the PBE exchange-correlation functional already incorporates mean-field correlation effects. An isolated monolayer at $U=0$ is itself a ferromagnetic insulator with a 0.14~eV gap and 1~$\mu_\text{B}$/SoD (Figure~S8), confirming that DFT captures substantial correlation effects. In multilayer systems, however, the increased bandwidth quenches magnetic order---only the 3L slab in the \textit{AL} configuration retains a residual moment of 0.4~$\mu_\text{B}$/SoD on the unpaired layer, while all other configurations (4L and above) are nonmagnetic, consistent with the absence of experimentally observed long-range magnetic ordering in 1T-TaSe$_2$ \cite{wilson1969}.

\textit{Metallic and insulating surface domains.} The coexistence of metallic and insulating domains at the 1T-TaSe$_2$ surface, observed by STM/STS \cite{chen2022b, zhang2022a} and ARPES \cite{straub2025, mignani2025, ren2025}, follows naturally from our energetics. The insulating domains correspond to the reconstructed \textit{A}-bilayer surfaces, i.e., the ground-state \textit{LLLLLLA} together with its nearly degenerate subsurface variants (\textit{LLLLLIA}, \textit{LLLLLHA}; Figure~S6), while the metallic domains are metastable bulk-terminated \textit{LLLLLLL} regions kinetically trapped during cleavage or cooling and possibly stabilized by local strain, rapid quenching, or impurity pinning. The $\sim$20~meV/SoD energy difference is large enough to drive reconstruction but small enough to permit coexistence. Minor contributions from \textit{LLLLLLM} or \textit{LLLLLLB} configurations ($\sim$12--16~meV/SoD above \textit{LLLLLLL}), which exhibit shallow pseudogaps (Figure~S3), may account for the intermediate small-gap ($\sim$0.1--0.2~eV) domains reported experimentally \cite{chen2022b, ren2025}. Two recent ARPES studies independently support this picture, attributing the insulating surface character to interlayer dimerization rather than Mott physics \cite{straub2025, mignani2025}.

\textit{Zero-bias features at liquid-helium temperature.} The STS data in Figure~2f were taken at liquid-nitrogen temperature; STS at liquid-helium temperature resolves additional features \cite{chen2022b}. The insulating surfaces divide into two kinds, without and with a small zero-bias peak (the latter termed weakly metallic), while the metallic surfaces (termed strongly metallic) exhibit a pronounced, narrow zero-bias peak reminiscent of the Kondo resonances observed in 1T/1H heterostructures \cite{vano2021, wan2023}. Our energetics naturally accounts for the two insulating kinds: they correspond to the presence or absence of a second bilayer beneath the surface bilayer, namely the fully gapped \textit{LLLLALA}-type and the \textit{LLLLLLA}-type surfaces (including their subsurface variants \textit{LLLLLIA} and \textit{LLLLLHA}), whose residual coupling to the metallic third layer leaves a weak zero-bias weight (Figure~2b,c); both are energetically accessible (Figure~2d). The sharp peak of the metallic surfaces instead points to a low-temperature Kondo resonance: their undimerized top layer retains a narrow half-filled band at $E_\text{F}$ (Figure~S6b) whose local moments---quenched in our calculations---apparently survive and are Kondo-screened by the metallic layers beneath. This scenario calls for temperature-dependent STS measurements and many-body calculations beyond the present mean-field treatment.

\textit{Thickness independence as a diagnostic.} The band-insulator gap is set by local bilayer hybridization and is therefore inherently insensitive to the total thickness, as confirmed from 2L to bulk by both our calculations and STS measurements (Figure~4) \cite{tian2024}; a Mott scenario would instead require fine-tuned cancellations of dimensionality and screening to reproduce the same gap at every thickness. Previous interpretations of insulating few-layer 1T-TaSe$_2$ as a dimensionality-driven Mott transition \cite{tian2024} can thus be reinterpreted within our framework: the insulating character arises from the same bilayer reconstruction that governs the bulk surface.

\textit{Trilayer films: exfoliated versus epitaxial.} Exfoliated trilayer samples exhibit the same $\sim$0.4~eV insulating gap as thicker films \cite{tian2024} (Figure~4b). In contrast, epitaxially grown trilayer films show a semimetallic spectrum with a much smaller peak-to-peak separation of $\sim$0.25~eV \cite{chen2020}. Our calculations reconcile this apparent contradiction: the two experiments probe the two inequivalent faces of the same ground-state trilayer---an \textit{A}-stacked bilayer plus a single layer. Exfoliation exposes the bilayer face with its hybridization gap, whereas in epitaxial films the interaction with the substrate can place the bilayer at the substrate side, exposing the single layer. The exposed single layer develops a magnetic moment (0.39, 0.89, and 1.00~$\mu_\text{B}$/SoD at $U=0$, 0.5, and 1.0~eV), and the spacing between its spin-split peaks grows linearly with $U$ (0.03, 0.10, and 0.15~eV for the same three values), producing a dip at $E_\text{F}$ flanked by two peaks without a full gap (Figure~S9), in agreement with the semimetallic trilayer spectra of Ref.~\citenum{chen2020}.

\textit{Universality of CDW surface reconstruction.} The analogy with 1T-TaS$_2$ provides a useful point of comparison. Despite different bulk ground states---\textit{AL} (band insulator) for TaS$_2$ and \textit{L} (metal) for TaSe$_2$---both surfaces converge on the same solution: an \textit{A}-interface bilayer that minimizes surface energy. The reconstruction transforms a Mott-insulating surface into a band insulator in TaS$_2$ and a metallic surface into a band insulator in TaSe$_2$, but in both cases the gap is opened by the same interlayer hybridization of Ta $5d_{z^2}$ orbitals. These observations suggest that CDW-driven surface reconstruction may arise more broadly in layered systems where interlayer CDW coupling is appreciable. In this sense, the surface should be viewed not as a passive termination of the bulk but as an active site of structural reorganization with significant electronic consequences.

\newpage

\section*{Methods}
\subsection*{Density functional theory calculations}

We performed density functional theory (DFT) calculations using the Vienna ab initio simulation package (VASP) \cite{VASP1, VASP2}. These calculations employed the projector-augmented-wave method \cite{PAW}, the generalized gradient approximation \cite{PBE}, and the DFT~+~$U$ scheme developed by Dudarev {\it et al.}\ \cite{Dudarev1998} with $U=0$ for Ta $5d$ orbitals. The choice of $U=0$ is motivated by the stronger interlayer coupling in 1T-TaSe$_2$ compared to 1T-TaS$_2$, which yields wider bandwidths and weaker effective correlations; this choice is validated by the excellent agreement between calculated and experimental spectra. For comparison, we also performed calculations for 1T-TaS$_2$ using $U=1.25$ eV, as established previously \cite{lee2023}. To test the robustness of our conclusions, selected 1T-TaSe$_2$ calculations were repeated at finite $U$ values (Figures~S5, S7, and S9). To account for van der Waals interactions, we used the Tkatchenko-Scheffler approach \cite{vdw-ts}. Spin-orbit interactions were not incorporated, as previous calculations determined their effects to be insignificant for the electronic states near the Fermi level in 1T-TaS$_2$ \cite{darancet2014, ritschel2018}. The electronic wave functions were expanded using a plane-wave basis set, with cutoff energies of 280 and 224 eV for the bulk and surface calculations of 1T-TaSe$_2$, respectively. The higher cutoff in the bulk calculations suppresses Pulay-stress errors during variable-cell optimization; for the fixed-cell slab calculations we verified that relative slab energies computed at the two cutoff levels agree within a few meV/SoD. We carried out $k$-space integration using a $4\times4\times8$ mesh in the Brillouin zone of the $\sqrt{13}\times\sqrt{13}\times1$ supercell. To optimize the lattice constants and atomic positions for each stacking configuration, we used the variable cell optimization method in VASP. For \textit{L} stacking, the calculated lattice constants were $a=3.47$~\AA\ and $c=6.18$~\AA, in good agreement with the experimental values of $a=3.48$~\AA\ and $c=6.26$~\AA\ \cite{brouwer1980} (deviations of $-0.3$\% and $-1.3$\%, respectively). To model the surface, we employed a periodic slab geometry, where each slab comprised two to eight TaSe$_2$ layers, and the thickness of the vacuum region exceeded $14$~\AA. We relaxed all atoms until all residual forces were less than 0.01 eV/\AA.

\providecommand{\latin}[1]{#1}
\makeatletter
\providecommand{\doi}
  {\begingroup\let\do\@makeother\dospecials
  \catcode`\{=1 \catcode`\}=2 \doi@aux}
\providecommand{\doi@aux}[1]{\endgroup\texttt{#1}}
\makeatother
\providecommand*\mcitethebibliography{\thebibliography}
\csname @ifundefined\endcsname{endmcitethebibliography}
  {\let\endmcitethebibliography\endthebibliography}{}

\newpage

\begin{figure*}[p]
	\includegraphics[scale=1.1]{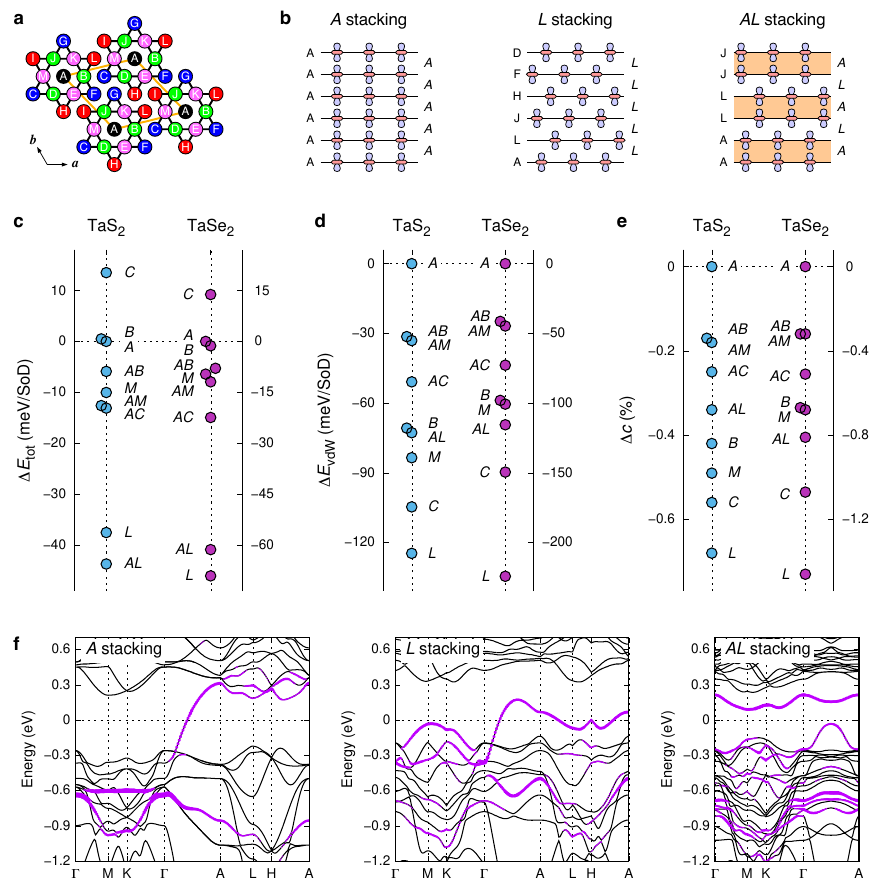}
	\caption{\textbf{CDW stacking order in bulk 1T-TaSe$_2$.}
		(a) Labeling convention for Ta atoms within a Star-of-David (SoD) cluster. Adapted with permission from ref~\citenum{lee2019}. Copyright 2019 American Physical Society.
		(b) Schematic illustration of the vertical arrangement of localized orbitals at the central Ta sites for the \textit{A}, \textit{L}, and \textit{AL} CDW stacking structures. The left markings indicate the SoD center positions in successive layers, and the right markings (italics) denote the CDW stacking interfaces.
		(c-e) Calculated relative total energies ($\Delta E_\text{tot}$), van der Waals energy contributions ($\Delta E_\text{vdW}$), and interlayer spacing changes ($\Delta c$) for various CDW stacking structures, referenced to the \textit{A} stacking. For comparison, results for 1T-TaS$_2$ are also shown.
		(f) Electronic band structures for three representative CDW stacking structures (\textit{A}, \textit{L}, and \textit{AL}). The violet intensity along each band represents the spectral weight of the central Ta $5d_{z^2}$ orbital within the SoD cluster.
	}
\end{figure*}

\begin{figure*}[p]
	\includegraphics[scale=1.1]{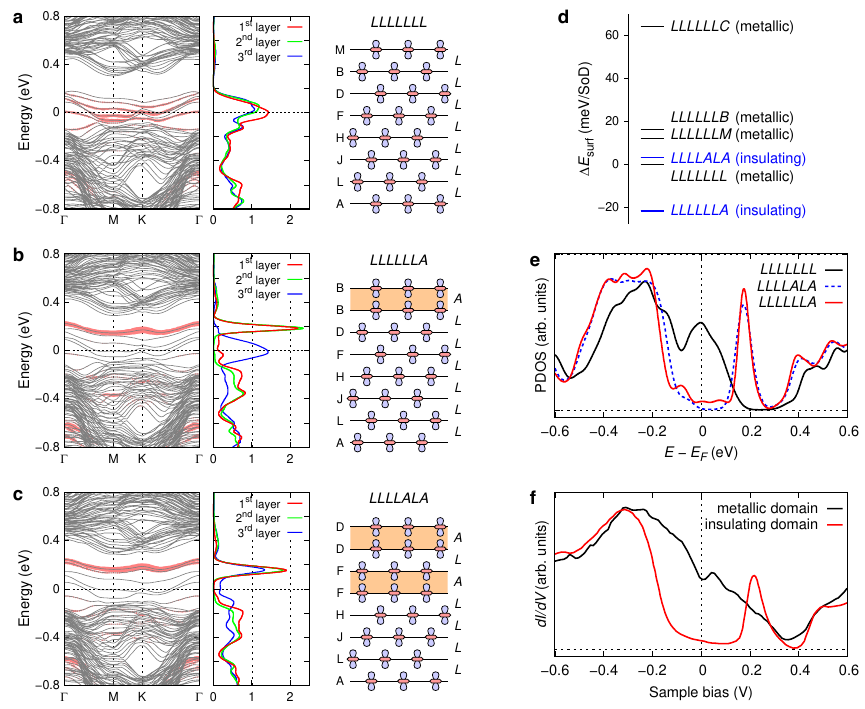}
	\caption{\textbf{Surface stacking--dependent electronic structures of 1T-TaSe$_2$.}
		(a--c) Left: Electronic band structures of 8-layer slabs with distinct surface CDW stacking at the upper surface, while the lower surface is fixed in the lowest-energy \textit{L} stacking. The red intensity along each band represents the spectral weight of the central Ta $5d_{z^2}$ orbital within the top-layer SoD cluster.
		Middle: Projected density of states (PDOS) of the central Ta $5d_{z^2}$ orbital for the first (red), second (green), and third (blue) layers from the surface.
		Right: Schematic illustration of the CDW stacking sequence, with the SoD center positions labeled on the left and the interface types on the right.
		(d) Relative surface energies ($\Delta E_\text{surf}$, in meV/SoD) of 8-layer slabs with different surface CDW stacking configurations, referenced to the \textit{LLLLLLL} surface (surface formation energy: 2.23~eV/SoD). Three representative cases are shown in (a--c); higher-energy cases are presented in Figure~S3.
		(e) PDOS of Ta $d_{z^2}$ orbitals summed over all 13 Ta atoms in the surface-layer SoD cluster for representative stacking motifs.
		(f) Differential conductance ($dI/dV$) from scanning tunneling spectroscopy (STS), measured at liquid-nitrogen temperature at two characteristic domains on the surface of bulk 1T-TaSe$_2$, showing metallic versus insulating behavior. Adapted with permission from ref~\citenum{zhang2022a}. Copyright 2022 American Physical Society.
	}
\end{figure*}

\begin{figure*}[p]
	\includegraphics[width=\textwidth]{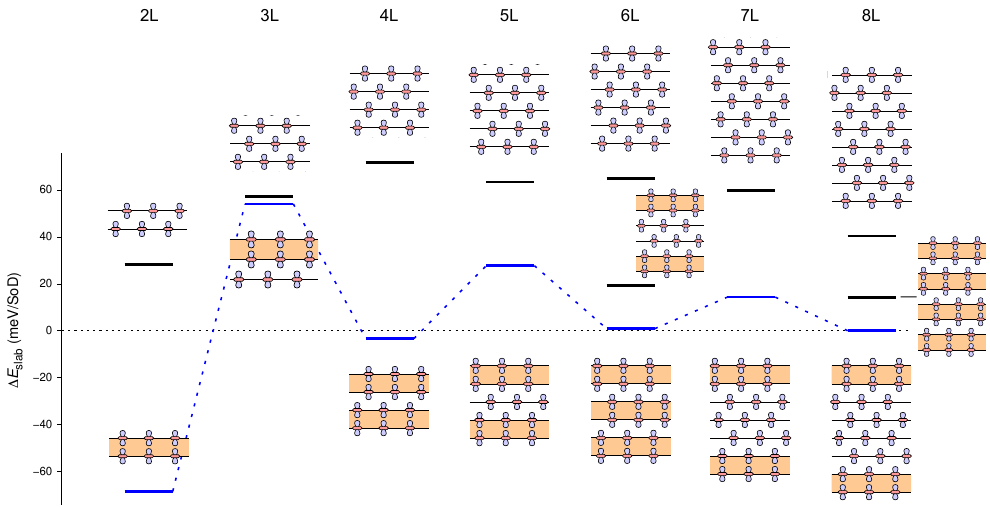}
	\caption{
		\textbf{Thickness-dependent energetics of 1T-TaSe$_2$ slabs.}
		Relative slab energies ($\Delta E_\text{slab}$, in meV/SoD) for 2--8 layer structures, referenced to the thick-slab limit of the ground-state \textit{ALLLLLA} configuration. For each thickness, the uniform \textit{L} stacking is compared with configurations in which one or both surfaces adopt an \textit{A}-interface bilayer (orange shading). The blue dashed line traces the ground-state energy as a function of thickness. At every thickness, the \textit{A}-interface bilayer termination is energetically preferred over the uniform \textit{L} stacking.
	}
\end{figure*}

\begin{figure*}[p]
	\includegraphics[scale=1.1]{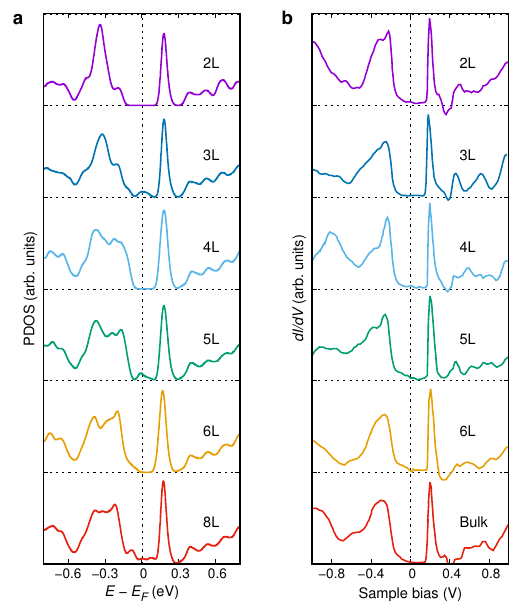}
	\caption{
		\textbf{Thickness-dependent surface electronic states of 1T-TaSe$_2$ slabs.}
		(a) Calculated PDOS of Ta $d_{z^2}$ orbitals summed over all 13 Ta atoms in the topmost-layer SoD cluster for the most stable slab configuration at each thickness (2--6 layers and bulk (8L)). In all cases, the surface is terminated by an \textit{A}-interface bilayer. The characteristic double-peak structure with a gap of $\sim$0.4~eV persists from 2L to 8L, confirming that the band-insulator character of the surface originates from the local bilayer hybridization and is independent of the total number of layers. Minor variations in peak width and position reflect changes in coupling to the underlying layers.
		(b) Differential conductance ($dI/dV$) spectra from scanning tunneling spectroscopy (STS) measured at the insulating domains on the surfaces of few-layer (2--6L) and bulk 1T-TaSe$_2$. The experimental spectra consistently exhibit the same insulating gap feature across all thicknesses, in excellent agreement with the calculated PDOS in (a). Adapted from ref~\citenum{tian2024}. Copyright 2024 The Author(s), published by Oxford University Press under a CC-BY license.
	}
\end{figure*}

\end{document}